\documentclass[prd,showpacs,twocolumn,aps]{revtex4-1}
\usepackage{slashed}
\usepackage{mathrsfs}
\usepackage{amsfonts}
\usepackage{amsmath}
\usepackage{amssymb}
\usepackage{revsymb}
\usepackage{graphicx}
\usepackage{mathrsfs}
\usepackage{bm}
\usepackage{bbm}
\usepackage{psfrag}
\usepackage{color}
\usepackage{hyperref}

\usepackage{natbib}

\usepackage[usenames,dvipsnames]{xcolor}

\newcommand{\ora}{\overrightarrow}
\newcommand{\ola}{\overleftarrow}
\newcommand{\oraS}{\ora{S_{fi}}}
\newcommand{\olaS}{\ola{S_{fi}}}

\newcommand{\be}{\begin{equation}} 
\newcommand{\ee}{\end{equation}} 
\newcommand{\bea}{\begin{eqnarray}} 
\newcommand{\eea}{\end{eqnarray}} 

\newcommand{\eps}{\varepsilon}
\newcommand{\mbf}[1]{\mathbf{#1}}
\newcommand{\trm}[1]{\textrm{#1}}

\newcommand{\figref}[1]{Fig. \ref{#1}}

\newcommand{\eqnref}[1]{Eq. (\ref{#1})}

\newcommand{\tr}{\trm{tr}\,}
\newcommand{\Ecr}{E_{\trm{cr}}}
\newcommand{\psibar}{\overline{\psi}}

\newcommand{\sk}{\slashed{\varkappa}}
\newcommand{\sA}{\slashed{A}}
\newcommand{\vkap}{\varkappa}
\newcommand{\vphi}{\varphi}

\newcommand{\Ai}{\trm{Ai}}

\newcommand{\xp}{x^{\prime}}
\newcommand{\kp}{k^{\prime}}

\definecolor{oorange}{rgb}{1.0,0.5,0.0}

\begin{document}
\title{Double Compton scattering in a constant crossed field}
\author{B. \surname{King}}
\email{ben.king@plymouth.ac.uk}
\affiliation{
    School of Computing and Mathematics, Plymouth University, 
    Plymouth PL4 8AA, UK}

\date{\today}
\begin{abstract}
Two-photon emission of an electron in an electromagnetic plane wave of vanishing 
frequency is calculated. The unpolarised probability is split into a two-step process, which 
is shown to be exactly equal to an integration over polarised subprocesses, and a one-step process, 
which is found to be dominant over the formation length. The assumptions of neglecting spin 
and simultaneous emission, commonly used in numerical simulations, are 
discussed in light of these results.
\end{abstract}
\maketitle

\section{Introduction}
It is well known that when an electron is accelerated by an electromagnetic field, it radiates 
\cite{jackson75}. When the wavelength of the radiation in the rest frame of the electron 
is much larger than the Compton wavelength, it is well 
described by classical electrodynamics. With ``photon emission'' we are referring to those 
situations 
in which shorter wavelengths are generated and a quantum electrodynamical description is 
necessary. Single-photon emission of electrons in a plane-wave background, commonly referred to as  
``[nonlinear] Compton scattering'', was first calculated five decades ago in monochromatic waves 
\cite{nikishov64, kibble64} and some time later, the effect of finite pulse-shapes 
\cite{narozhny96,boca09,harvey09,mackenroth10,heinzl10,mackenroth11}, electron spin 
\cite{krajewska13}, photon \cite{king13a} and external-field polarisation 
\cite{ivanov04,bashmakov14} have been 
investigated. Single Compton scattering has been experimentally observed in the weakly nonlinear 
regime \cite{bula96,bamber99} and advances in 
laser technology have motivated studying two-photon 
emission in a pulsed plane wave background with possible experimental signatures having been 
discussed in the literature \cite{seipt12,mackenroth13} (a review of strong-field QED effects can be 
found in \cite{ritus85,marklund06,dipiazza12}).
\newline

In the current paper, we will calculate two-photon emission of an 
electron in an electromagnetic plane wave of vanishing frequency, the so-called ``constant crossed 
field''. On the one hand, this will 
allow us to produce the first in-depth analysis of two-photon emission in the non-perturbative 
region of large quantum non-linearity parameter. On the other, the constant crossed 
field background is the one used overwhelmingly in current numerical simulations that combine 
particle-in-cell 
propagation of charged particles with Monte Carlo generation of ``quantum'' events such as photon 
emission \cite{sokolov10,nerush11,elkina11,blackburn14,green14,green14b,mironov14}. Such 
simulations iterate single-vertex processes to approximate higher-order ones and our double-vertex 
calculation can assess the 
faithfulness of this approximation. 
In their calculation of the two-loop 
electron mass operator in a constant crossed field \cite{morozov75}, Morozov and Ritus have also 
produced expressions for 
the total probability of two-photon emission. However, their emphasis on infra-red 
behaviour and the brevity of exposition differ significantly from the intention of the present
article.

\section{Polarised single Compton scattering}
\begin{figure}[!h]
\noindent\centering
\hspace{0.025\linewidth}
\includegraphics[draft=false]{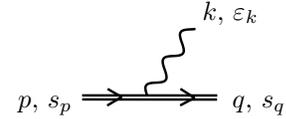}
\hfill
\caption{The Feynman diagram for single Compton scattering.}
\label{fig:single_compton_feynman} 
\end{figure}

We begin by calculating the probability for single Compton scattering taking into account the 
polarisation of all incoming and outgoing particles. Although our analysis is for electrons, 
analogous arguments apply to positrons. We consider the process:
\bea
e^{-} \to e^{-} + \gamma
\eea
in a constant field background where $e^{-}$ refers to an electron and 
$\gamma$ to a photon. The corresponding Feynman diagram is given in 
\figref{fig:single_compton_feynman} where we note $p$ and $q$ are the incoming and outgoing 
electron's four-momentum respectively and $s_{p}$ and $s_{q}$ the corresponding spin four-vectors, 
with $k$ and $\eps_{k}$ the emitted photon's four-vector and polarisation four-vector where 
$s_{p}^{2}=s_{q}^{2}=\eps_{k}^{2}=-1$.
The scattering matrix for single-photon emission can be written as \cite{mandl10}
\bea
S_{fi}= 
ie\int\!d^{4}x~\overline{\psi}_{q}(x)\frac{\sqrt{4\pi}\slashed{\eps}_{k}^{\ast}\mbox{e}^{ik\cdot 
x}}{\sqrt { 2k^{ 0 } V } }
\psi_{p}(x),
\eea
where $\slashed{v} = \gamma\cdot v$ for a four-vector $v$, $\gamma$ are the gamma matrices 
\cite{mandl10} and $e>0$ is the elemental positron charge and $m$ its mass, with $V$ the 
normalisation volume. The 
wavefunction for an electron in a plane wave electromagnetic background is given by the Volkov 
solution to the Dirac equation \cite{volkov35}:
\bea
\psi_{p}(x) &=& \Big[1+\frac{e \sk\sA[\vphi(x)]}{2(\varkappa\cdot p)}\Big] 
\frac{u_{p}}{\sqrt{2p^{0}V}} 
\mbox{e}^{iS(p,\varphi(x))}\\
S(p,\varphi) &=& -px - \int^{\varphi}_{0} d\phi \,\left[\frac{e\,p\cdot A(\phi)}{\vkap\cdot p} - 
\frac{e^{2}A^{2}(\phi)}{2(\vkap\cdot p)}\right],
\label{eqn:Volkov}
\eea
where the external-field phase $\vphi=\vkap\cdot x$ for external-field wavevector $\vkap$ and 
vector potential $A^{\mu}(\vphi)$ and the normalisation of the electron spinor $u_{p}$ will be 
explained shortly. The probability for polarised single-photon emission $P_{\gamma}$ is then 
given by 
\bea
P_{\gamma} = 
V^{2}\int\,\frac{d^{3}k}{(2\pi)^{3}}\,\frac{d^{3}q}{(2\pi)^{3}}~\tr|S_{fi}|^{2}. 
\label{eqn:Pgam1}
\eea
When squaring the trace of $|S_{fi}|^{2}$, the electron spin is introduced using a standard 
method of writing the spin density matrix for electron momentum $p$ as \cite{landau4}
\bea
u_{p}\overline{u}_{p} = \frac{1}{2}(\slashed{p} + m)(1-\gamma^{5}\slashed{s}_{p}).
\eea
To calculate $P_{\gamma}$, we employ the method developed by
Nikishov and Ritus (see \cite{ritus85}, and a more detailed application to single Compton 
scattering in \cite{king13a}). A key part of this method is that after momentum conservation has 
been applied to the integration of one outgoing particle's momentum in \eqnref{eqn:Pgam1}, the 
remaining integrand is independent of the projection of remaining outgoing particle momenta on the 
vector potential. As there exists a one-to-one map from these momentum projections to the 
value 
of the external field phase, this divergent integral is crucially re-interpreted as an integral 
over phase, for example:
\bea
\int d(k\cdot a^{(1)})\bigg|_{\trm{on shell}} = \int\frac{d\vphi}{J},
\eea
where the prescription ``on shell'' refers to momentum conservation having been applied to the 
integral in $q$ and $J=\big|\partial \vphi/\partial (k\cdot a^{(1)})\big|$ is the Jacobian, which 
is independent of 
$\vphi$. When selecting 
a basis for the electron spin and photon polarisation vectors, 
it 
is advantageous to maintain this symmetry so that the Nikishov-Ritus method can be consistently 
applied. For polarised photons but unpolarised electrons, a basis in which 
$\eps_{k}\vkap=0$ is sufficient to preserve the symmetry in momentum. When electron spin is also 
included, a basis in which $s_{p}\vkap=0$ greatly simplifies expressions, but is insufficient alone
to preserve the momentum symmetry. A key difference between electron spin and photon polarisation 
is that the precession of the former due to the external field is already included to all orders by 
using the dressed Volkov 
propagator whereas the evolution of the latter, described by a dressed photon propagator, is a 
higher-order effect in $\alpha=e^{2}\approx1/137$, the fine structure constant, the first 
corrections 
of which enter at $O(\alpha^{2})$ \cite{dinu14a,dinu14b,king10b,king14a}. Since single Compton 
scattering 
is to first order in $\alpha$, 
the evolution of photon polarisation does not enter the calculation. Conversely, the evolution of 
the electron spin \emph{can} be described by invoking the Bargmann-Telegedi-Michel equation 
\cite{bargmann59}, 
\bea
\frac{ds_{p}}{d\tau} = \frac{e}{m}\left[\frac{g}{2}\,F\cdot 
s_{p}+\left(\frac{g}{2}-1\right)\frac{\left(s_{p}\cdot F\cdot p\right) p}{m^{2}}\right],
\eea
where $F$ is the Faraday tensor \cite{jackson75}, $g\approx2$ is the electron's gyromagnetic ratio 
\cite{gabrielse06,gabrielse07} and $\tau$ is the proper time. By choosing a spin basis such that 
$ds_{p}/d\tau=0$ as well as $s_{p}\vkap =0$, the momentum symmetry appears and ensures the  
Nikishov-Ritus method can be applied automatically.
\newline

Our choice of vector potential, polarisation basis and spin basis that allow the 
Nikishov-Ritus method to be straightforwardly applied is
\bea
A(\vphi) &=& a^{(1)}g^{(1)}(\vphi) + a^{(2)}g^{(2)}(\vphi)\\
\eps^{(1,2)}_{k} &=& a^{(1,2)} - \frac{k\cdot a^{(1,2)}}{k\cdot\varkappa} ~ \varkappa,\\
\zeta_{p} &=& a^{(2)} - \frac{p\cdot a^{(2)}}{p\cdot \varkappa} ~ \varkappa,
\eea
where $a^{(1)}\cdot a^{(2)} = 0$, $a^{(1)}\cdot a^{(1)} = a^{(2)}\cdot a^{(2)} = 
-1$ and $g^{(2)}(x) = 0$. That this choice ensures the spin basis $\zeta_{p}$ does not 
precess, can be seen from $d\zeta_{p}/d\tau = 0$ or in the rest frame of the electron, 
$\pmb{\zeta}_{p}\wedge\mbf{B}=\mbf{0}$, where $\mbf{B}$ is the external magnetic field 
vector.\newline

We now specify the calculation to a constant crossed field background by choosing $g^{(1)}(\vphi) = 
(m\xi/e)\varphi$ where $\xi=e|p\cdot F|/m|\vkap\cdot p|$ is the classical non-linearity 
parameter \cite{ilderton09}, which can be written as $\xi = m \mathcal{E}/\vkap^{0}$, with 
$\mathcal{E}=E/\Ecr$ the ratio of the electric field amplitude $E$ to the critical field 
$\Ecr=m^{2}/e$. In a constant field, $\xi$ is formally infinite as the limit 
$\vkap^{0}\to0$ is 
taken in expressions for the total rate. In a constant crossed field background, 
total rates are 
functions of another gauge- and relativistic invariant referred to as each particle's quantum 
non-linearity parameter, which for a momentum $p$ is given by $\chi_{p}= e|p\cdot F|/m^{3}$. 
Recognising $\xi\vphi$ as being independent 
of 
the limit $\vkap^{0}\to0$, let us define the rate of single Compton scattering per unit 
normalised external field phase $R_{\gamma} = P_{\gamma}/\xi\int d\varphi$. Suppose we 
expand the polarisation and spins as:
\begin{gather}
\eps_{k} = c_{1}\eps^{(1)}_{k} + c_{2}\eps^{(2)}_{k}\\
s_{p} = \sigma_{p} \zeta_{p}\qquad s_{q} = \sigma_{q} \zeta_{q}
\end{gather}
where $c_{1,2}\in\{0,1\}$, $\sigma_{p,q}\in\{-1,0,1\}$, we then find
\bea
R_{\gamma} = -\frac{\alpha}{\chi_{p}^{2}}\!\int_{0}^{\chi_{p}}\!\!d\chi_{k}\left[C^{\cdot} 
\Ai(z)+C' 
\Ai'(z) + C_{1}\Ai_{1}(z)\right],\nonumber \\ \label{eqn:Rg}
\eea
where $\Ai(x)=\frac{1}{\pi}\int_{0}^{\infty}\!dk\,\cos(kx+k^{3}/3)$ is the Airy function, $\Ai'(x)$ 
its derivative, $\Ai_{1}(x) = \int_{x}^{\infty}\!dk~\Ai(k)$ and
\bea
z &=& \left[\frac{\chi_{k}}{\chi_{p}(\chi_{p}-\chi_{k})}\right]^{2/3}
\eea
\bea
C^{\cdot} &=& 
z\left[(\sigma_{p}+\sigma_{q})(\chi_{k}-2c_{k}^{2}\chi_{p})-2(1-c_{k}^{2})\sigma_{q}\chi_{k} 
\right]\nonumber\\
C' &=& 
\chi_{k}z^{1/2}\left(1-c_{\sigma}c_{\delta}\sigma_{p}\sigma_{q}\right)+\frac{2}{z}
\left(1+\sigma_{p}\sigma_{q}\right)\left(1-c_{\sigma}c_{\delta}\right)\nonumber\\
C_{1} 
&=&1+\sigma_{p}\sigma_{q}-\frac{c_{\sigma}c_{\delta}\sigma_{p}\sigma_{q}\chi_{k}^{2}}{\chi_{p}(\chi_
{
p}-\chi_{k})},\nonumber\\\label{eqn:Cgs}
\eea
where the single polarisation parameter $c_{k} = c_{1} = \sqrt{1-c_{2}^{2}}$ has been introduced 
with $2c_{\sigma}=c_{1}+c_{2}$ and $c_{\delta} = c_{2}-c_{1}$ simplifying notation. The rate for 
single Compton scattering with unpolarised electrons but a polarised photon \cite{king13a} can be 
recovered if one takes the limit in \eqnref{eqn:Rg} of zero spin $\sigma_{p},\sigma_{q}\to 0$. If 
one then averages over photon polarisations, the totally unpolarised rate \cite{nikishov64} is 
acquired. We note the appearance of an $\Ai(\cdot)$ function in the integrand of $R_{\gamma}$, 
which is only 
present if electron spin is taken into account and is completely absent in standard 
unpolarised calculations. This term will appear again in the double Compton scattering calculation. 
\newline

\begin{figure}[!]
\noindent\centering
\hspace{0.025\linewidth}
\includegraphics[draft=false,width=5.5cm]{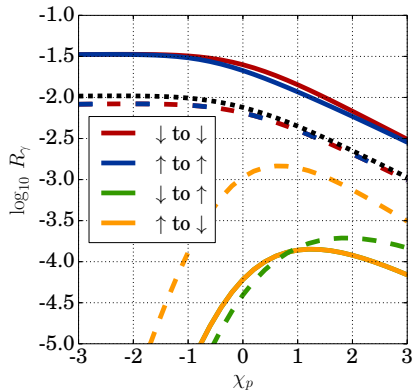}
\hfill
\caption{The rate of polarised single Compton scattering 
$R_{\gamma}$. The solid (dashed) lines refer to a photon scattered into the 
$\eps_{k}^{(1)}$ ($\eps_{k}^{(2)}$) polarisation state and the dotted line refers to the 
unpolarised rate.}
\label{fig:Igtotplot} 
\end{figure}

The rate of single Compton scattering for different combinations 
of initial and final polarisations of 
particles is illustrated in \figref{fig:Igtotplot}, where solid (dashed) lines correspond to 
emitted photons in 
a polarisation state $\eps_{k}=\eps_{k}^{(1)}$ ($\eps_{k}=\eps_{k}^{(2)}$) and $\uparrow$ 
($\downarrow$) to spin states $\sigma=1$ ($\sigma=-1$). 
We notice that in this non-precessing spin basis, the ``no-flip'' polarisation channels in which 
the spin of the electron 
is unchanged after photon emission are in general favoured more than the ``spin-flip'' 
channels, which are 
suppressed, particularly for small $\chi_{p}$.

\section{Double Compton scattering}
\begin{figure}[!h]
\noindent\centering
\hspace{0.025\linewidth}
\includegraphics[draft=false]{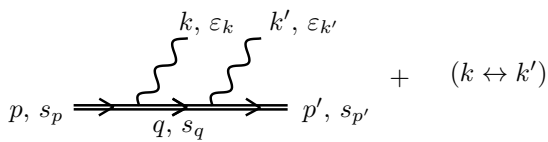}
\hfill
\caption{The Feynman diagram for double Compton scattering. The prescription on the exchange 
term $\left(k\leftrightarrow k'\right)$ also applies to swapping the polarisation vectors.}
\label{fig:double_compton_feynman} 
\end{figure}

Let us now turn to the calculation of double photon emission by an electron in a plane wave field. 
We are considering the process
\bea
e^{-} \to e^{-} + \gamma + \gamma'
\eea
and the corresponding Feynman diagram is given in \figref{fig:double_compton_feynman}. The 
transition matrix element for this process is
\begin{widetext}
\bea
S_{fi} &=& \oraS+\olaS\\
\oraS &=& -e^{2}\int\!d^{4}\xp\,d^{4}x~\psibar_{p^{\prime}}(\xp) 
\frac{\sqrt{4\pi}\slashed{\eps}^{\prime\,\ast}_{\kp}}{\sqrt{2Vk^{\prime\,0}}}\mbox{e}^{i\kp\xp}
G(\xp,x)\mbox{e}^{i k x} 
\frac{\sqrt{4\pi}\slashed{\eps}^{\ast}_{k}}{\sqrt{2Vk^{0}}}\psi_{p}(x), \label{eqn:Sfi2}\\
G(\xp,x) &=& 
\int\!\frac{d^{4}q}{(2\pi)^{4}}\left[1+\frac{e\slashed{\vkap}\slashed{A}(\vphi')}{2\,\vkap\cdot 
q}\right]\mbox{e}^{
iS(q,\vphi^{\prime})} 
\frac{\slashed{q}+m}{q^{2}-m^{2}+i0}\mbox{e}^{-iS(q,\vphi)}\left[1+\frac{e\slashed{A}(\vphi) 
\slashed{\vkap}
} { 2\,\vkap\cdot q}\right]
\eea 
\end{widetext}
where $\oraS$ corresponds to the first diagram in \figref{fig:double_compton_feynman} and 
$\olaS$ to making the replacements $k,\eps_{k}\leftrightarrow k',\eps_{k'}$.
The method of calculation is similar to in the previous section, with the added complication of 
having an extra diagram due to bosonic exchange symmetry of identical outgoing photons as well as 
a fermionic propagator (a detailed example of the Nikishov-Ritus approach being 
applied to a two-vertex process is given in the calculation of electron-seeded pair creation in 
\cite{king13b}). We will confine ourselves to calculating the unpolarised double Compton scattering 
probability $P_{\gamma\gamma}$
\bea
P_{\gamma\gamma} = \frac{1}{4}V^{3}\int 
\frac{d^{3}p'}{(2\pi)^{3}}\,\frac{d^{3}k}{(2\pi)^{3}} 
\frac{d^{3}k'}{(2\pi)^{3}}~\tr|S_{fi}|^{2}, \label{eqn:Pgg}
\eea
with the pre-factor including an average over initial electron spins and symmetry factor 
due to identical diagrams. The modulus squared amplitude contains each exchange term 
mod-squared plus interference terms
\bea
|S_{fi}|^{2} = |\oraS|^{2}+|\olaS|^{2}+\olaS\oraS^{\ast}+\oraS\olaS^{\ast}. \label{eqn:Sfigg}
\eea
Now suppose the $p'$ integral in \eqnref{eqn:Pgg} is performed by evaluating the 
standard total momentum-conserving delta function that arises in scattering matrix calculations and 
that the integral over $k$ and $k'$ remain. In the standard fashion, these integrals can be 
re-interpreted as
\bea
\int d(k\cdot a^{(1)})\,d(k'\cdot a^{(1)})\bigg|_{\trm{on shell}} = \int 
\frac{d\vphi_{+}\,d\vphi_{-}}{J}.
\eea
Here, $\vphi_{+}=(\vphi_{x'}+\vphi_{x})/2$ is the average of the stationary phases in the function 
describing the probability of photon emission at spacetime points $x$ and $x'$ and hence 
corresponds to the centre in phase between two emissions and
$\vphi_{-} = \vphi_{x'}-\vphi_{x}>0$ is the phase the electron travels between emissions, where $J$ 
is the corresponding Jacobian. Part of the 
integrand is completely independent of both these phases, and we term this the \emph{two-step} 
process, part of the integrand depends only on $\vphi_{-}$, and we term this the 
\emph{one-step} process and the remaining part of the integrand depends on both phases. Since these 
phase integrations are formally infinite in a constant crossed field, the part that depends on both 
phases and gives a finite answer, will be dropped from the calculation. It can be shown 
\cite{king13b} that this neglected part corresponds to the interference terms in 
\eqnref{eqn:Sfigg}. As the total probability includes an integration over 
both photon momenta, one can then replace $|S_{fi}|^{2} \to 2|\oraS|^{2}$ in the integrand. If we 
again define the rate $R_{\gamma\gamma} = P_{\gamma\gamma}/\xi\int d\vphi_{+}$ where 
$\vphi=\vphi_{x}$ is the phase of the first emission the total rate becomes
\bea
R_{\gamma\gamma} = R^{(2)}_{\gamma\gamma} + R^{(1)}_{\gamma\gamma}
\eea
where the superscripts indicate the two- and one- step rates accordingly and
\bea
R^{(2)}_{\gamma\gamma} &=& \mathcal{I}^{(2)}_{\gamma\gamma}~\xi\int 
d\vphi_{-}\label{eqn:R2}\\
R^{(1)}_{\gamma\gamma} &=& \mathcal{I}^{(1)}_{\gamma\gamma}\label{eqn:R1},
\eea
where the quantities $\mathcal{I}_{\gamma\gamma}$ are free of divergences associated with the 
infinite expanse of the background and will be referred to as the \emph{dynamical} part of the 
rate in 
contrast to the \emph{spacetime} factors that multiply them in the probability, such as 
integrations over external-field phase.

\subsection{Two-step double Compton scattering}
The two-step process actually comprises terms from both the on- and off- shell part of the fermion
propagator, where off-shell terms are essential to preserve causality \cite{king13b}, so it is not 
synonymous with the ``on-shell'' part. We find 
\bea
\mathcal{I}_{\gamma\gamma}^{(2)} &=& 
-\alpha^{2}\int\!d\chi_{k}d\chi_{k'}\,\left[C^{\cdot\cdot}\Ai(z)\Ai(z')+\right.\nonumber \\
 &&\left. C^{\prime\prime}\Ai'(z)\Ai'(z')+C^{\prime1}\Ai'(z)\Ai_{1}(z'
)+\nonumber \right.\\
&& \left. 
C^{1\prime}\Ai_{1}(z)\Ai'(z')+C^{11}\Ai_{1}(z)\Ai_{1}(z') \right],
\eea
where we have defined
\bea
z &=& \left[\frac{\chi_{k}}{\chi_{p}\chi_{q}}\right]^{2/3}, \quad z' = 
\left[\frac{\chi_{k'}}{\chi_{p'}\chi_{q}}\right]^{2/3};
\eea
\bea
C^{\cdot\cdot} &=& -\frac{zz'\chi_{p}\chi_{p'}}{(\chi_{p}\chi_{q})^{2}}\\
C^{\prime\prime} &=& 
\frac{1}{(\chi_{p}\chi_{q})^{2}}\left(\frac{2}{z'}+\chi_{k'}z^{\prime\,1/2}\right)\left(\frac{2}{z
}
+\chi_{k}z^{1/2}\right)\\
C^{\prime1} &=& \frac{1}{(\chi_{p}\chi_{q})^{2}}\left(\frac{2}{z}+\chi_{k}z^{1/2}\right)\\
C^{1\prime} &=& \frac{1}{(\chi_{p}\chi_{q})^{2}}\left(\frac{2}{z'}+\chi_{k'}z^{\prime\,1/2}\right)
\\
C^{11}  &=& \frac{1}{(\chi_{p}\chi_{q})^{2}},
\eea
with $\chi_{q}=\chi_{p}-\chi_{k}$ and $\chi_{p'}=\chi_{p}-\chi_{k}-\chi_{k'}$. Written in this way, 
upon comparison with the 
probability for unpolarised single photon scattering (averaging over photon polarisation and setting 
$\sigma_{p}, \sigma_{q} \to 0$ in \eqnref{eqn:Cgs}), one can see that the two-step process is the 
integral over the product of single Compton scattering. We find that the unpolarised two-step rate 
can indeed be exactly factorised in 
terms of single Compton scattering processes, when the intermediate electron's spin is taken 
into account and assumed in an unchanged state when the second photon is emitted (we note here the 
relevance of choosing a non-precessing spin basis)
\bea
R_{\gamma\gamma}^{(2)} = \frac{1}{2}\sum_{\sigma_{q}} \int d\chi_{q} 
\frac{\partial R_{\gamma}(\chi_{p})}{\partial \chi_{q}}\, R_{\gamma}(\chi_{q})\,\frac{\xi}{2}\int 
d\vphi_{-}. \label{eqn:Rgg2fac}
\eea

\subsection{One-step double Compton scattering}
The one-step process involves an extra integration over a variable related to the virtuality of 
the propagating electron.
\bea
\mathcal{I}_{\gamma\gamma}^{(1)} =-\frac{\alpha^{2}}{4\pi}\int\!d\chi_{k}\,d\chi_{k'}\,dt 
~\frac{\mathcal{A}(t)+\mathcal{A}(-t)-2\mathcal{A}(0)}{t^{2}} \label{eqn:I1}
\eea
\bea
\mathcal{A}(t) &=& C^{\cdot\cdot}_{t}\Ai(z_{t})\Ai(z_{t}')+ 
\nonumber \\
&&C^{\prime\prime}_{t}\Ai'(z_{t})\Ai'(z_{t}')+C^{\prime1}_{t}\Ai'(z_{t})\Ai_{1}(z_{t}')+\nonumber 
\\&&C_{t} ^ { 1\prime }\Ai_{1}(z_{t})\Ai'(z'_{t})+C^{11}_{t}\Ai_{1}(z_{t})\Ai_{1}(z'_{t})
\eea
\bea
z_{t} &=& z+\frac{t}{z^{1/2}}, \qquad z'_{t} = 
z'-\frac{t}{z'^{1/2}},
\eea
where we have defined
\bea
\frac{C^{\cdot\cdot}_{t}}{C^{\cdot\cdot}} &=& 1-\frac{t}{2} 
\frac{\chi_{q}(\chi_{p}+\chi_{q})(\chi_{q}+\chi_{p'})}{\chi_{k}\chi_{k'}}\\
\frac{C^{\prime\prime}_{t}}{C^{\prime\prime}} &=& 1\\
\frac{C^{\prime1}_{t}}{C^{\prime1}} &=& 1+\frac{t\chi_{k}}{2}\\
\frac{C^{1\prime}_{t}}{C^{1\prime}} &=&1-\frac{t\chi_{k'}}{2}\\
\frac{C^{11}_{t}}{C^{11}} &=& 
1+t\frac{\chi_{q}^{2}-\chi_{k}\chi_{k'}}{2\chi_{q}}+\frac{t^{2}}{4}(\chi_
{ p }\chi_{p'}-\chi_{k}\chi_{k'}).
\eea
After performing the integral over the propagator variable $t$ and $\chi_{k'}$, the remaining 
integral in $\chi_{k}$ diverges $\sim 1/\chi_{k}$ for $\chi_{k}\to 0$. This well-known 
infra-red divergence was reported in other calculations in double Compton scattering 
\cite{loetstedt09a,loetstedt09b,seipt12}, and should be cancelled when self-energy corrections are 
included \cite{morozov75}.
\newline

It can be shown \cite{king13b} that the one-step process can be written as a term originating 
from the interference between two- and one- step parts of the amplitude plus a term originating 
solely from the one-step part of the amplitude. Here, as in electron-seeded pair creation in a 
constant crossed field, parts of the one-step probability are negative, however since the phase 
factor multiplying the two-step term is formally divergent, the total probability is 
non-negative. Unlike for electron-seeded pair creation, we find no threshold value of $\chi_{p}$ 
above which the one-step process becomes positive.

\section{Approximations used in simulation}
We turn now to a comparison of the analytical results for double Compton scattering with 
approximations used in numerical simulations. In particular, we investigate two assumptions. 
First, the neglecting of electron spin, which we found essential to correctly factorising 
the 
two-step part of double Compton scattering and which produced new terms in the integrand. Second, 
the neglecting of the one-step process also known as ``simultaneous'' two-photon emission 
by an electron. 

\subsection{Electron spin}
As an example of employing the constant crossed field approximation, the unpolarised probability of
the two-step process in a slowly-varying external field can be written as a double iteration of 
single photon emission:
\bea
P_{\trm{ccf}}^{(2)} = \frac{1}{2}\sum_{\sigma_{q}} \int
d\vphi_{\xi'}d\vphi_{\xi} d\chi_{q}\,P_{\gamma}[\chi_{q}]\,
\frac { \partial P_{\gamma}[\chi_{p},\chi_{q}]}{\partial\chi_{q}}, \label{eqn:CCF1}
\eea
where we have allowed the external field to depend on the phase by defining $\vphi_{\xi} = \vphi\, 
\xi(\vphi)$, $\vphi_{\xi'} = \vphi'\, 
\xi(\vphi')$ and $\vphi_{\xi'}>\vphi_{\xi}$,  
$\chi_{p}=\chi_{p}(\vphi_{\xi})$, $\chi_{q}=\chi_{q}(\vphi_{\xi'})$ and $\chi_{q}>\chi_{p}$. If the 
external field is taken to be exactly constant $\xi(\vphi) = \xi$, \eqnref{eqn:CCF1} leads back to 
\eqnref{eqn:Rgg2fac}. It is therefore consistent to 
only include single Compton scattering in a simulational approach, and allow for the process to 
occur multiple times, as is performed in numerical simulation. To investigate the importance of 
including electron spin, we plot the dynamical part of the two-step double Compton scattering 
$\mathcal{I}_{\gamma\gamma}^{(2)}$ using \eqnref{eqn:Rgg2fac}, as well as the case when the 
propagating electron is forced to be unpolarised by setting $\sigma_{q}\to0$ in 
\eqnref{eqn:Rgg2fac}, which we label $\overline{\mathcal{I}}_{\gamma\gamma}^{(2)}$. For 
comparative purposes, we also plot 
Morozov and Ritus' asymptotic limits for the two-step process by introducing the function
\be
\mathcal{I}_{\trm{MR}}^{(2)}(\chi_{p}) =\left\{
\begin{array}{cl}
 \frac{25\alpha^{2}}{12} & \chi_{p}<1\\
 \frac{119 \alpha^{2}}{270}\Gamma(\frac{1}{3})(\frac{3}{\chi_{p}})^{2/3} & 
\chi_{p}>1
\end{array}
\right.,
\ee
where $\chi_{p}<1$ ($\chi_{p}>1$) is the $\chi_{p} \ll 1$ ($\chi_{p} \gg 1$) asymptotic limit.
\begin{figure}[!h]
\noindent\centering
\hspace{0.025\linewidth}
\includegraphics[draft=false,width=5cm]{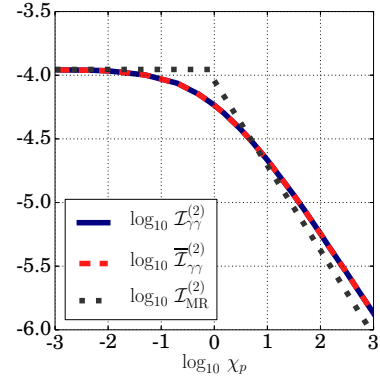}
\hfill
\caption{The dynamical part of the two-step probability $\mathcal{I}_{\gamma\gamma}^{(2)}$ with the 
approximation of using unpolarised electrons $\overline{\mathcal{I}}_{\gamma\gamma}^{(2)}$ and the 
asymptotic limits of Morozov and Ritus $\mathcal{I}^{(2)}_{\trm{MR}}$}
\label{fig:I2tot} 
\end{figure}
In \figref{fig:I2tot}, we note that the intermediate electron's spin seems to make very little 
difference to the total probability for double Compton scattering. We highlight the property of 
Compton scattering in a constant crossed field, that $\partial P_{\gamma}/\partial \chi_{k} \sim 
\chi_{k}^{-2/3}$ for $\chi_{k}\to 0$. Although the differential probability diverges as 
$\chi_{k}\to 0$, the total probability remains finite (the softening of this well-known infra-red 
divergence has 
recently been studied in \cite{lavelle06,dinu12,ilderton13a}). How to take into account this 
divergent 
number of photons is handled in a variety of ways by numerical simulations, but often a hard energy 
or $\chi_{k}$ cutoff is introduced, below which the effects of the emitted radiation are included 
using the classical equations of electrodynamics. By introducing a 
cutoff in our analysis, for example, neglecting photons with $\chi_{k}<0.1$ (several orders of 
magnitude larger than what is usually considered \cite{harvey14}), the effect of 
the electron spin to the total probability is still only at the few percent level. Therefore, it 
seems that treating the propagating particle as a scalar in numerical simulations of multi-photon 
emission of electrons in intense laser pulses is a consistent approximation.

\subsection{Simultaneous photon emission}
\begin{figure}[!h]
\noindent\centering
\hspace{0.025\linewidth}
\includegraphics[draft=false,width=5cm]{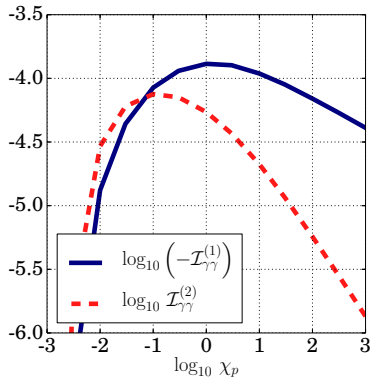}
\hfill
\caption{The absolute dynamical part of the one-step probability 
$\mathcal{I}_{\gamma\gamma}^{(1)}$ is 
compared to the two-step probability $\mathcal{I}_{\gamma\gamma}^{(2)}$ including an infra-red 
cutoff $\Delta_{k}=\Delta_{k'}=10^{-5}$. 
}
\label{fig:I1totplotC} 
\end{figure}

With simultaneous photon emission, we are referring to the one-step process given by the integral 
in \eqnref{eqn:I1}. We have already commented that this is divergent and negative. Therefore 
comparison of the factorised two-step process with the ``rest'' of the probability of two-photon 
emission is not possible at $O(\alpha^{2})$ without including self-energy terms. Although 
it makes little sense to compare two- and one- step processes without including self-energy 
terms, as they may also appreciably affect photon emission for large $\chi$ values, we 
\emph{can} assess how much of the one-step process is neglected in simulation codes when a 
cutoff in the emitted photons' $\chi$ parameter is used. For example, in 
the simulation of photon emission by a single electron in \cite{harvey14}, a cutoff of 
$\chi_{k}=\Delta_{k}=10^{-5}$ was chosen. In \figref{fig:I1totplotC} we compare the total 
probability of the one- and two- step processes using $\chi_{k}$ and $\chi_{k'}$ cutoffs of 
$\Delta_{k}=\Delta_{k'}=\Delta=10^{-5}$.
For $\chi_{p}\gg 1$, $\mathcal{I}_{\gamma\gamma}^{(1)}$ agrees with the scaling and sign given by 
Morozov and Ritus for this limit, although we were unable to compare results in a quantitative 
way. 
We note that the absolute ratio of dynamical parts of the one- to the two- step probabilities 
becomes 
larger than unity already at $\chi_{p}\approx 0.1$, and linearly increases to more than an order of 
magnitude for $\chi_{p}\gtrsim 50$. After repeating the calculation for 
a range of cutoffs $\Delta\in[10^{-7},10^{-2}]$, although the exact ratio is weakly 
cutoff-dependent, for $\chi_{p}\in[1,1000]$, the linear increase in the ratio appears 
cutoff-independent.
\newline

Over the formation length $1/\xi\varkappa^{0}$ the one-step process can clearly be dominant 
compared to 
the two-step process, which implies current numerical approaches are inappropriate for 
simulating double photon emission in this parameter regime. We recall the total rate for 
two-photon emission in a constant 
crossed field is given by
\bea
R_{\gamma\gamma} = \mathcal{I}^{(2)}_{\gamma\gamma}~\frac{\xi}{2}\int d\vphi_{-} 
+ \mathcal{I}^{(1)}_{\gamma\gamma}+\frac{\mathcal{I}^{(0)}_{\gamma\gamma}}{\xi\int d\vphi_{-}} ,
\eea
where $\mathcal{I}^{(0)}_{\gamma\gamma}$ is the interference between exchange terms, neglected as 
the phase factor is formally infinite, but reintroduced here when discussing approximating more 
complicated fields as constant crossed. The constant crossed approximation to 
an arbitrary field is expected to be valid when $\xi\gg 1$ and $\chi_{p}$ is much larger than the 
two electromagnetic invariants $e^{2}F^{2}/m^{4}$ and $e^{2}FF^{\ast}/m^{4}$ for Faraday tensor $F$ 
and its dual $F^{\ast}$ \cite{ritus85}. The present results imply that the further assumption made 
in simulations 
that the rate of generating $n$ photons is due to the $n$-step process, is questionable for 
$n=2$ when $\xi \lesssim O(10^{2})$ and $\chi_{p}\gtrsim 1$. To demonstrate this, we use the 
constant crossed field approximation in \eqnref{eqn:CCF1} and 
\bea
P^{(1)}_{\trm{ccf}} = \int\!d\vphi_{\xi}\, 
R_{\gamma\gamma}^{(1)}[\chi_{p}(\vphi_{\xi})]
\eea
to estimate double Compton scattering in the field of 
a laser pulse $E = E_{0}\,\cos^{4}(\pi\vphi/2\Phi)\cos\vphi$ for pulse width $\Phi = 
\vkap^{0}\tau$.
\begin{figure}[!h]
\noindent\centering
\hspace{0.025\linewidth}
\includegraphics[draft=false,width=5cm]{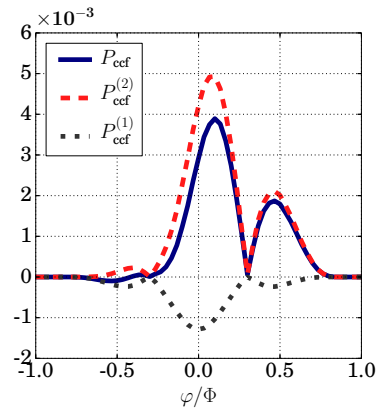}
\hfill
\caption{For a few-cycle pulse with field strength equivalent to $500\,\trm{TeV}$ focussed to 
a focal width of $10\,\mu\trm{m}$ of $910\,\trm{nm}$ wavelength with peak $\xi=10$, and pulse 
duration $5\,\trm{fs}$ counterpropagating with $10\,\trm{GeV}$ seed electrons.}
\label{fig:LCFAplot} 
\end{figure}
In \figref{fig:LCFAplot}, the relative difference in photon yield due to including the one-step 
process is $-25\%$, however the main qualitative difference due to
including the one-step process is the instant when two-photon emission starts to become 
significant, which is predicted to occur a half-cycle later. 
\newline

Although we have seen that one requires background fields less than two orders of magnitude 
larger than the formation length $1/\xi\vkap^{0}$ and $\chi_{p} \gg 0.1$ for the one-step 
process to 
be comparable to the two-step one, the analysis raises the question of how accurate it is 
to simulate an $n$-photon emission including just the $n$-step process.


\section{Conclusion}
Using a non-precessing spin basis to describe electron polarisation, the probability of 
a spin flip following single Compton scattering in a constant crossed field was found to 
be suppressed. The unpolarised rate for double Compton scattering in a constant crossed field was 
written as a sum of a two-step process, which is exactly factorisable as single Compton 
scattering integrated over the longitudinal momentum of a polarised intermediate electron, and a 
one-step process, which dominates the total probability over the formation length 
$1/\xi\vkap^{0}=\lambdabar \Ecr/E$. Regarding numerical simulation of double photon emission, we 
found the assumption that the 
intermediate electron is unpolarised, to be accurate to the few percent level, depending on the 
photon energy cutoff used and that although simultaneous two-photon emission can be 
much more probable than sequential emission, when the external field's spatial dimensions are more 
than around two orders of magnitude larger than the single-photon formation length, the sequential 
process is dominant. The availability of intense ultra-short laser pulses would allow measurement 
of the quantum interference between simultaneous and sequential production channels in the 
nonlinear domain.

\section{Acknowledgments}
B.K. acknowledges the stimulating and enlightening discussions with A. Fedotov and M. Legkov that 
inspired this project and conversations on the infra-red with M. Lavelle and A. Ilderton. This work 
was in part 
supported by the Russian Fund for Basic Research, grant No.13-02-90912.

\bibliography{current}
\end{document}